\documentclass[%
 reprint,
 amsmath,amssymb,
 aps,
]{revtex4-2}

\usepackage{graphicx}
\usepackage{dcolumn}
\usepackage{bm}
\usepackage{textcomp}
\usepackage{graphicx}
\usepackage{float}
\usepackage{here}
\usepackage{svg,ulem}
\usepackage{hyperref,color}
\usepackage[english]{babel}

\begin{document}

\preprint{APS/123-QED}

\title{Compressed sensing quantum state tomography for qudits{: A comparison of Gell-Mann and Heisenberg-Weyl {observable} bases}}

\author{Yoshiyuki Kakihara}
 \email{chys24001@g.nihon-u.ac.jp}
\author{Daisuke Yamamoto}
\email{yamamoto.daisuke21@nihon-u.ac.jp }
\author{Giacomo Marmorini}
\email{marmorini.giacomo@nihon-u.ac.jp}
\affiliation{
Department of Physics, College of Humanities and Sciences, Nihon University, {Sakurajosui, Setagaya,} Tokyo 156-8550, Japan
}

\date{\today}

\begin{abstract}
 Quantum state tomography (QST) is an essential technique for reconstructing the density matrix of an unknown quantum state from measurement data, {crucial for} quantum information processing. However, conventional QST requires an exponentially growing number of measurements {as the system dimension increases, posing a significant challenge} for high-dimensional systems. To {mitigate} this issue, compressed sensing quantum state tomography (CS-QST) has been proposed, {significantly reducing} the required number of measurements. In this study, we investigate the impact of basis selection in CS-QST {for qudit systems, which are fundamental to} high-dimensional quantum information processing. Specifically, we compare the efficiency of the generalized Gell-Mann (GGM) and Heisenberg-Weyl {observable (HWO)} bases by numerically reconstructing density matrices and evaluating reconstruction accuracy using fidelity and trace distance metrics. Our results demonstrate that, while both bases allow for successful density matrix reconstruction, the HWO basis becomes more efficient as the qudit dimension increases. Furthermore, we {find the best} fitting curves that estimate the number of measurement operators required to achieve a fidelity of at least 95\%. These findings highlight the significance of basis selection in CS-QST and provide valuable insights for optimizing measurement strategies in high-dimensional quantum state tomography.
\end{abstract}

\maketitle


\section{\label{sec:level1}Introduction}
With the {rapid} development of quantum {processors, quantum} circuits, and {experimental platforms}, understanding the quantum states they {generate} has become a crucial research topic in quantum information science. {A key} tool {for this purpose} is quantum state tomography (QST), which {reconstructs the density matrix of an unknown quantum state by measuring expectation values of a complete set of operators}. QST has been {employed} to characterize quantum states in experimental platforms such as trapped ions~\cite{ion}, solid-state qubits~\cite{solidqbit}, and photonic {systems}~\cite{superconducting}. However, QST faces a scalability challenge: the number of required {measurement} operators grows exponentially with the number of qubits $n$. One {approach} to {address } this {challenge} is compressed sensing quantum state tomography (CS-QST){,} proposed by D. Gross $et$ ${al}$ \cite{CS}. They {demonstrated that for {a multi-qubit} system, the density matrix can be approximately reconstructed} using the singular value thresholding (SVT) algorithm~\cite{SVT}, {requiring only $O(rd\log^2d)$ measurement bases instead of the original $d^2$}, where $r$ and $d$ represent the rank and dimension of the matrix, respectively. CS-QST has since been implemented in {photonic~\cite{CSexp} and ion-trap~\cite{CSexpion} systems}.

{While a qubit is a two-level quantum unit analogous to a classical bit (but capable of existing in a superposition of states), a qudit extends this concept to a system with $k$ levels, offering advantages in information storage and processing while potentially reducing} the complexity of quantum circuits~\cite{Qudit}. {Qudits} have been realized in various physical {platforms,} such as {photons}~\cite{babazadeh-17,lu-20} superconducting transmon{s}~\cite{Quditexp}, trapped ion{s}~\cite{trappedion2022,hrmo-23,meth-25}, {molecular nanomagnets}~\cite{chicco-23}, {ultracold atoms}~\cite{lindon-23} and 
{nuclear magnetic resonance systems}~\cite{gedik-15}. {There has also been growing interest in developing quantum algorithms tailored for qudit systems~\cite{qutrit_algo_by_transmon,transmon_algo,quditalgo}, exploring their potential in high-dimensional quantum computing.} QST for qudit{s~\cite{quditQST,CSQD}  faces a scalability challenge} similar to {that} of qubit{s: the number of required measurement operators grows} exponentially with the number of qudits. 
{Additionally, for a qudit of dimension $k$, the number of {one-body} operators increases as $k^2$. }

{Naturally, QST for qudits requires that the matrix representation of measurement operators has the appropriate dimension; for instance a single qudit with $k=3$, or  qutrit, requires a  matrix representation of dimension three. The choice of operator basis is not unique and may be motivated by theoretical and experimental considerations.
While the standard choice in power-of-two dimension is the Pauli operator basis, two common bases for qudit tomography of arbitrary $k$-dimensional systems are the generalized Gell-Mann (GGM)~\cite{GGM} and Heisenberg-Weyl observable (HWO)~\cite{HW} bases, which have also been used experimentally in {Ref.}~\cite{GGMexpe2011} (GGM) and {Ref.}~\cite{HWexpe2020} (HW{O}). }
{A key distinction} between these  bases is given by {their} coherence {properties, as} defined in {Ref.}~\cite{cohe}. {Coherence properties underlie the mathematical proof of quantum compressed sensing  (later refined in terms of the ``restricted isometry property''\cite{flammia-12,kalev-15}).} {In the GGM basis, the coherence values grow with $k$, owing to the diagonal operators in particular, whereas in the HWO basis they are independent of $k$ (and all elements share the same spectral norm); consequently, only the latter rigorously meet the hypothesis of the CS-QST theorem in Ref.~\cite{cohe}. Additionally, Asadian $et$ $al.$~\cite{HW} noted that all HW observables have maximal rank, which should make them more ``universal'', {\it i.e.,} capable of encoding a certain amount of information characterizing any kind of quantum states}.
{In this context, it is worth comparing how different choices of basis affect the efficiency of practical implementations of CS-QST of higher-dimensional density matrices in qudit systems; this ought to be seen as a complementary analysis to the mathematically rigorous results of Ref.~\cite{cohe}. Crucially, providing quantitative insights into the above question can guide future QST experiments, given that setting up the measurement of ``theoretically ideal'' observables may be extremely costly or outright impossible, and has to be weighed against using an alternative, easier to implement, observable basis.}

{In this paper, w}e compare the efficiency of {the} GGM and HWO bases in CS-QST to investigate the impact of coherence differences on matrix reconstruction. Specifically, we {calculate} the fidelity and trace distance of {the} reconstructed density matrices to determine which basis enables more accurate reconstruction with fewer measurements {and evaluate that distinction quantitatively.} {Furthermore, we determine the minimum number of measurement operators required to ensure that} the fidelity {(minus one standard deviation)} {remains above} 95$\%$ and {examine how this number varies with system parameters}. {From a} different perspective, we  also compare the efficiency {of} CS-QST {when applied to} density matri{ces} of the same dimension, {but making use of different qudit dimensions $k$ and basis choices}. As a specific example, {for a} density matrix {of dimension $d=16$, we perform CS-QST} using the Pauli basis {for} a 4-qubit system, the GGM or HWO basis {for a system of two $k=4$ qudits, and a single $k=16$ qudit}. 

Based on these numerical analyses, we find that {for low qudit dimensions}, there is no significant difference in reconstruction efficiency between the GGM and HWO bases. However, as the qudit dimension increases, the HWO basis enables more efficient density matrix reconstruction. {That said}, the difference remains small, and {while} the accuracy is lower {with} the GGM basis, the density matrix can still be reconstructed {with sufficient precision}. {Additionally, we observe that CS-QST using the HWO basis achieves the same level of efficiency as the Pauli basis for quantum states of the same dimension, regardless of the qudit dimension $k$. However, when the GGM basis is used, the reconstruction efficiency deteriorates as $k$ increases. These findings provide valuable {indications for an optimal basis selection in CS-QST future experiments, aimed at an efficient and feasible}  density matrix reconstruction in high-dimensional quantum systems.}

{The remainder of this paper is organized as follows. In Sec.~\ref{QST}, we review the general framework of QST for qudits. Section~\ref{QSTA} introduces the basic formulation of QST, while Sec.~\ref{QSTB} presents two typical bases used for qudit systems: the GGM basis and the HWO basis. In Sec.~\ref{sec_coherence}, we discuss the coherence properties of these bases, which are relevant to the performance of CS-QST. Section~\ref{sec3} presents our numerical simulations. In Sec.~\ref{Ginibre ensemble}, we explain how random test states are generated using the Ginibre ensemble. Section~\ref{methods} describes the simulation setup, including noise modeling and the reconstruction algorithm. In Sec.~\ref{results}, we present the main results, focusing on comparisons between the GGM and HWO bases for CS-QST applied to both two-qudit systems and systems with a fixed Hilbert space dimension. Finally, Sec.~\ref{summary} provides a summary of our findings.}

\section{Quantum state tomography for qudits}
\label{QST}

\subsection{Quantum state tomography}
\label{QSTA}

{QST is a technique used to} reconstruct the density matrix $\rho$ of an unknown quantum state {by making measurements on an ensemble of identical systems}. {For a system of $N$ qubits, the density matrix $\rho$ is represented by a $2^N \times 2^N$ matrix. Due to their orthogonality, the Pauli {operators} form a natural basis for expanding the multi-qubit quantum state $\rho$. {Specifically}, $\rho$ can generally be written as follows:} 
\begin{widetext}
\begin{eqnarray}
    \rho=\frac{1}{2^{N}}\sum_{i_1=0}^3\sum_{i_2=0}^3\cdots\sum_{i_{N}=0}^3\langle\sigma_{i_1}\otimes\sigma_{i_2}\otimes\cdots\otimes\sigma_{i_{N}}\rangle ~\sigma_{i_1}\otimes\sigma_{i_2}\otimes\cdots\otimes\sigma_{i_{N}},
\end{eqnarray}
\end{widetext}
{where $\sigma_0$ is the identity matrix, and $\sigma_1,\sigma_2,\sigma_3$ correspond to the standard Pauli matrices. The {normalization} factor $1/2^N$ ensures that ${\rm Tr}[\rho]=1$. Therefore, to {completely} reconstruct $\rho$, one needs to measure the expectation values of the Pauli  operators $\sigma_{i_1}\otimes\sigma_{i_2}\otimes\cdots\otimes\sigma_{i_N}$, excluding the identity  $(\sigma_{0})^{\otimes N}$. However, the number of distinct Pauli operators  grows exponentially with $N$, {namely} as $4^{N}-1$, {which implies} that an exponentially increasing number of measurements is required to obtain all the necessary expectation values. Additionally,}   {to obtain  each expectation value,  the quantum state must be repeatedly prepared and undergo the same quantum measurement; only collecting a sufficient number of outcomes in this way the expectation value can be accurately estimated.}

{The above protocol for QST can be generalized to qudits with arbitrary dimensions $k$ by using a certain orthogonal  basis, which is given by the product of {single-qudit} operators $\omega_{i_n}$, as follows:}
\begin{widetext}
\begin{eqnarray}
    \rho={\frac{1}{k^N}}\sum_{i_1=0}^{k^2-1}\sum_{i_2=0}^{k^2-1}\cdots\sum_{i_{N}=0}^{k^2-1}\langle \omega_{i_1}\otimes\omega_{i_2}\otimes\cdots\otimes\omega_{i_{N}}\rangle~\omega_{i_1}\otimes\omega_{i_2}\otimes\cdots\otimes\omega_{i_{N}}{\equiv\sum_{a=0}^{d^{2}-1}\langle B_a\rangle~B_a~~(d=k^N)},
    \label{quditQST}
\end{eqnarray}
\end{widetext}
{where $B_a$ corresponds to $(\omega_{i_1}\otimes\omega_{i_2}\otimes\cdots\otimes\omega_{i_N})/\sqrt{k^N}$ with $a$ being the result of converting  {$i_1i_2\dots i_N$} from base $k$ to decimal. Thus, the number of required measurement operators is $d^2-1$, which grows exponentially as $N$ increases, with a base of $k^2$. As for the choice of $\omega_{i}$ consisting of an orthonormal basis $\lbrace B_a\rbrace^{d^2-1}_{a=0}$, we focus on the two typical choices described in the next subsection.}

\subsection{Two {typical} bases for {qudit systems}} \label{QSTB}
\subsubsection{Generalized Gell-Mann basis}
{The generalized Gell-Mann (GGM) basis~\cite{GGM},}  {is formed by the standard set of generators of the SU($k$) Lie algebra}. 
The {elements of the} GGM basis are divided into three types as follows: 
\begin{description}
    \item[i)] Symmetric matrices
    
    \begin{align}
        \Lambda_s^{jj'}=|j\rangle \langle j'|+|j'\rangle \langle j |,\quad (0 \leq j < j'\leq k-1)
        \label{eq:GGM}
    \end{align}
    
    \item[ii)] Anti-symmetric matrices
    \begin{align}
        \Lambda_a^{jj'}=-i |j\rangle \langle j'|+i|j'\rangle \langle j |,\quad (0 \leq j < j'\leq k-1)
        \label{eq:GGM2}
    \end{align}
    \item[iii)] Diagonal matrices

    \begin{align}
        \begin{split}
        \Lambda_d^l=&\sqrt{\frac{2}{(l+1)(l+2)}}\Big(\sum_{j=0}^l|j\rangle\langle j|-(l+1)|l+1\rangle\langle l+1|\Big),\\[3mm] &( 0\leq l\leq k-2),
         \label{eq:GGM3}
        \end{split}
    \end{align}    
\end{description}
{where $j, j', l$ are integers, and the {basis states are} denoted as $|0\rangle, |1\rangle,\dots |k-1\rangle$. There are therefore $\frac{k(k-1)}{2}$ symmetric, $\frac{k(k-1)}{2}$ anti-symmetric, and $(k-1)$ diagonal elements. In total, we have $k^2-1$ distinct operators $\lambda_i\equiv \{\Lambda_s^{jj'},\Lambda_a^{jj'},\Lambda_d^l\}$ $(i=1,2,\dots,k^2-1)$, which satisfy the orthogonality relation}
\begin{align}
{\mathrm{Tr} [{ \lambda_i \lambda_{i'} }] = 2 \delta_{i,i'}.}
\end{align}

{For $k=2$ (qubit), the GGM matrices reduce to  the three Pauli matrices. For $k=3$ (qutrit)}, they are the usual Gell-Mann matrices \cite{gell-mann-62} defined as:
\begin{align}
\begin{split}
    &\lambda_1=
    \begin{pmatrix}
        0&1&0\\1&0&0\\0&0&0
    \end{pmatrix},\quad
    \lambda_2=
    \begin{pmatrix}
        0&-i&0\\i&0&0\\0&0&0
    \end{pmatrix},\\
    &\lambda_3=
    \begin{pmatrix}
        1&0&0\\0&-1&0\\0&0&0
    \end{pmatrix},\quad
    \lambda_4=
    \begin{pmatrix}
        0&0&1\\0&0&0\\1&0&0
    \end{pmatrix},\\
    &\lambda_5=
    \begin{pmatrix}
        0&0&-i\\0&0&0\\i&0&0
    \end{pmatrix},\quad
    \lambda_6=
    \begin{pmatrix}
        0&0&0\\0&0&1\\0&1&0
    \end{pmatrix},\\
    &\lambda_7=
    \begin{pmatrix}
        0&0&0\\0&0&-i\\0&i&0
    \end{pmatrix},\quad
    \lambda_8=\frac{1}{\sqrt{3}}
    \begin{pmatrix}
        1&0&0\\0&1&0\\0&0&-2
    \end{pmatrix}.
\end{split}
\label{eq:gell}
\end{align}
Appendix \ref{aappendixGGM} also provides the $k=4$ (quartit) case explicitly.

{Note that we employ the $k^2-1$ GGM matrices $\lambda_i$ $(i=1,2,\dots,k^2-1)$ renormalized by a factor of $\sqrt{k/2}$, along with the $k$-dimensional identity matrix, as the $k^2$  $\omega_i$'s in Eq.~\eqref{quditQST}. The renormalization factor $\sqrt{k/2}$ is introduced for convenience, ensuring that}
\begin{align}
{\mathrm{Tr} [{ \omega_i \omega_{i'} }] = k \delta_{i,i'}.}
\end{align}

\subsubsection{Heisenberg-Weyl {observable }basis}
{The matrices known as ``Heisenberg-Weyl (HW) matrices'' or ``Sylvester's generalized Pauli matrices'' are a non-Hermitian extension} of the Pauli matrices to higher dimensions~{\cite{sylvester-82}}. {Let us consider the ``shift operator'' $\mathcal{S}$ and ``clock operator'' $\mathcal{C}$, which act on the state bases} as $\mathcal{S}|j\rangle=|j+1\hspace{5pt} \mathrm{mod} \hspace{3pt} k\rangle$ and $\mathcal{C}|j\rangle =e^{i2\pi j/k}|j\rangle$. Using these operators, {the HW  matrices are defined by~\cite{HW}} 
\begin{align}
    {\mathcal{W}(l,m)=\mathcal{C}^l\mathcal{S}^me^{-i\pi lm/k}~(l,m=0,1,\dots, k-1),}\label{HWop1}
\end{align}
{which are unitary but not Hermitian}.

{In Ref.~\cite{HW}, Asadian ${\mathit{et}}$ ${\mathit{al.}}$ constructed a  basis of Hermitian single-qudit operators by taking linear combinations of the non-Hermitian HW matrices:}
\begin{align}
{W(l,m)=\chi\mathcal{W}(l,m)+\chi^*\mathcal{W}^\dagger(l,m),}\label{HWop2}
\end{align}
where $\chi=(1\pm i)/2$ {(in this paper, we choose the positive sign). These HW ``observable'' operators are Hermitian and satisfy the orthogonality relation}
\begin{align}
    \mathrm{Tr}[ {W}(l,m){W}(l',m')] =k \delta_{l,l'}\delta_{m,m'}.
\end{align}

{For $k=2$ (qubit), the HW observable operators reduce to the Pauli matrices together with the identity. For $k=3$ (qutrit), the specific matrix representations are given by}
\begin{widetext}
\begin{align}
\begin{split}
    &W(0,0)=
    \begin{pmatrix}
        1&0&0\\0&1&0\\0&0&1
    \end{pmatrix},\quad
    W(0,1)=
    \begin{pmatrix}
        0&\chi^*&\chi\\\chi&0&\chi^*\\\chi^*&{\chi}&{0}
    \end{pmatrix},\quad
    W(0,2)=
    \begin{pmatrix}
        0&\chi&\chi^*\\\chi^*&0&\chi\\\chi&\chi^*&0
    \end{pmatrix}, \\
    &W(1,0)=
    \begin{pmatrix}
        \chi+\chi^*&0&0\\0&\chi \theta+\chi^*\theta^*&0\\0&0&\chi \theta^*+\chi^*\theta
    \end{pmatrix},\quad
    W(1,1)=
    \begin{pmatrix}
        0&-\chi^*\theta&-\chi\theta\\-\chi\theta^*&0&-\chi^*\\-\chi^*\theta^*&-\chi&0
    \end{pmatrix},\quad
    W(1,2)=
    \begin{pmatrix}
        0&\chi\theta^*&\chi^*\theta^*\\\chi^*\theta&0&\chi\\\chi\theta&\chi^*&0
    \end{pmatrix},\\
    &W(2,0)=
    \begin{pmatrix}
    \chi+\chi^*&0&0\\0&\chi\theta^*+\chi^*\theta&0\\0&0&\chi\theta+\chi^*\theta^*
    \end{pmatrix},\quad
    W(2,1)=
    \begin{pmatrix}
        0&\chi^*\theta^*&\chi\theta^*\\ {\chi} \theta&0& {\chi^*} \\\chi^*\theta&\chi&0
    \end{pmatrix},\quad
    W(2,2)=
    \begin{pmatrix}
        0&\chi\theta&\chi^*\theta\\\chi^*\theta^*&0&\chi\\\chi\theta^*&\chi^*&0
    \end{pmatrix},
\end{split}
\end{align}
\end{widetext}
where $\theta$ denotes $e^{2\pi i/3}$. {The $k=4$ (quartit) case is provided explicitly in appendix \ref{aappendixGGM}.}

{We employ the $k^2$ HW observable operators (including the identity matrix $W(0,0)$) as the  $\omega_i$'s in Eq.~\eqref{quditQST}. We refer to this basis as the HW-observable basis, or simply, the HWO basis.}


\subsection{{C}oherence}\label{sec_coherence}
{In compressed sensing, the ease of recovery is influenced by the coherence between the target matrix and the basis matrices. A lower coherence typically leads to better recovery performance, as it implies that the target signal can be sparsely represented with fewer measurements, making it easier to reconstruct accurately using compressed sensing techniques. Here, we examine the GGM and HWO bases from the perspective of coherence, using the definition provided in Ref.~\cite{cohe}, to understand their impact on recovery performance.}

In Ref.~~\cite{cohe}, it is defined that a rank-$r$ $d\times d$ matrix $\rho$ to be recovered has coherence $\nu$ with respect to an orthonormal basis $\lbrace B_a\rbrace^{d^2-1}_{a=0}$ if either of the following holds:
\begin{itemize}
    \item[(i)]
\begin{align}
   \max_a\|B_a\|^2\leq\nu\frac{1}{d};
    \label{cohere}
\end{align}
\item[(ii)] the two estimates
\begin{align}
    \max_a\|\mathcal{P}_TB_a\|_2^2\leq2\nu\frac{r}{d}, \label{cohere2}\\
    \max_a{\rm Tr}[B_a \mathrm{sgn}(\rho)]^2\leq\nu\frac{r}{d^2}. \label{cohere3}
\end{align}
\end{itemize}
where $\|\sigma\|$ and $\|\sigma\|_2$ refer to the
spectral norm and Frobenius norm, respectively, of a matrix $\sigma$. Here $\mathcal{P}_T$ is the map
\begin{align}
    \mathcal{P}_T:\sigma \mapsto P_U\sigma +\sigma P_U-P_U\sigma P_U,
\end{align}
where $P_U$ is the orthogonal projection onto $U={\rm range}(\rho)$~\cite{cohe}. The diagonal matrix ${\rm sgn}(\rho)\equiv {\rm diag}[{\rm sgn}(\epsilon_1),{\rm sgn}(\epsilon_2),\dots,{\rm sgn}(\epsilon_d)]$, where $\epsilon_n$ is the $n$-th eigenvalue of $\rho$ and ${\rm sgn}(x)$ is the sign function with ${\rm sgn}(0)=0$.
While the condition in Eq.\eqref{cohere} depends solely on the properties of the basis $B_a$ itself, the conditions in Eqs.\eqref{cohere2} and~\eqref{cohere3} reflect the compatibility between the target state to be reconstructed and the measurement basis. Therefore, here we focus on the former to discuss the choice of basis {in general, having in mind an ensemble of random quantum states as target}. {We name the coherence value saturating Eq.~\eqref{cohere} the ``minimum coherence'' and denote it by}
\begin{align}
\nu_{\rm min}\equiv d\max_a\|B_a\|^2. \label{mincoherence}
\end{align}
It characterizes the worst-case concentration of {tomographic} weight across the basis elements and serves as a basis-dependent indicator of recoverability in compressed sensing.  
\color{black}

\begin{figure}[tb]
    \begin{center}
    \includegraphics[width=0.9\linewidth]{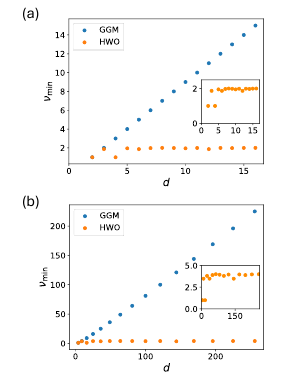}
    \end{center}
    \caption{{The minimum coherence $\nu_{\rm min}$, defined by Eq.~\eqref{mincoherence}, is plotted as a function of the dimension $d=k^N$, comparing the GGM and HWO bases. Panels (a) and (b) correspond to $N=1$ and $N=2$, respectively, with qudit dimensions ranging from $k=2$ to $k=16$. The insets show magnified views of the HWO results in each panel. Note that for $k=2$, both bases reduce to the Pauli basis.}}
    \label{fig:cohe}
\end{figure}

{In Figs.~\ref{fig:cohe}(a) and \ref{fig:cohe}(b), we plot the values of $\nu_{\rm min}$ for the cases of $N=1$ and $N=2$ {qudits}, respectively, comparing the GGM and HWO bases as a function of the dimension $d=k^N$. Due to the multiplicative property of the spectral norm with respect to the tensor product, $\nu_{\rm min}$ for $N$ {qudits} is simply given by $d \left(\max_i \|\omega_i\|^2/k\right)^N=\left(\max_i \|\omega_i\|^2\right)^N$. It can be seen that while $\nu_{\rm min}$ increases approximately linearly with $d$ (specifically $\nu_{\rm min}=(k-1)^N\sim d$) for the GGM basis, it remains small 
and nearly constant regardless of $k$ for the HWO basis. The specific values for $k\leq 16$ are summarized in Table~\ref{tableHW}. Interestingly, in particular, the case of $k=4$ for the HWO basis yields $\nu_{\rm min}=1$, indicating that all measurement operators are unitary, as in the Pauli ($k=2$) basis. For other values of $k$, the minimum coherence takes the form $\nu_{\rm min}\approx 2^N$, and becomes exactly $2^N$ when $k$ is a multiple of 8 (see Appendix~\ref{appendix2}). }
{These observations} suggest that, compared to the HWO basis, the number of operators required for accurate density matrix reconstruction {tends to} increase {with} the qudit dimension {when using} the GGM basis. 
\begin{table}[tb]
\caption{{Numerical values of the minimum coherence $\nu_{\rm min}$ for the HWO basis, evaluated for qudit dimensions up to $k=16$ in the case of $N=1$. For general $N$, the values can be obtained by raising the $N=1$ value to the power of $N$.}} 
\begin{center}
\label{tableHW}
\begin{tabular}{c|ccccccc}
     \hline\hline
     $k$ & 3 & 4 & 5 & 6 & 7 & 8 & 9 \\
     \hline
     $\nu_{\rm min}(N=1)$ & 1.866 & ~~1~~ & 1.951 & 1.866 & 1.975 & ~~2~~ & 1.985 \\
     \hline
\end{tabular}

\vspace{1ex}

\begin{tabular}{c|ccccccc}
     \hline
     $k$ & 10 & 11 & 12 & 13 & 14 & 15 & 16 \\
     \hline
     $\nu_{\rm min}(N=1)$ & 1.951 & 1.990 & 1.866 & 1.993 & 1.975 & 1.995 & ~~2~~ \\
     \hline\hline
\end{tabular}
\end{center}
\end{table}

{Moreover, we find that while the {spectral norms} $\|\omega_i\|^2$ remain independent of the index $i$ in the HWO basis, except for those operators that include the identity matrix, they vary depending on the specific operators chosen in the GGM basis. Specifically, the diagonal matrices (except for the one with $l=0$ in Eq.~\eqref{eq:GGM3}) have different $\|\omega_i\|^2$ values compared to the the off-diagonal ones, with the maximum value ($\max_i \|\omega_i\|^2=k-1$) attained by $\sqrt{k/2}\Lambda^{k-2}$. This implies that the information quality in the GGM basis can fluctuate depending on the choice of operators, potentially making accurate reconstruction more difficult. }

{These expectations are examined numerically in the next section by testing } the impact of using the GGM and HWO bases in CS-QST under {the following} two conditions:
(i) {p}erforming CS-QST with {a randomly selected set of} measurement operators to analyze how the choice of basis affects reconstruction accuracy{;}
(ii) {e}xamining how increasing the qudit dimension influences the efficiency of density matrix reconstruction.

{\section{Numerical simulations}}\label{sec3}
\subsection{Ginibre {ensemble} }
\label{Ginibre ensemble}
{As target states of the numerical tests of CS-QST, w}e generate a random density matrix $\rho$ {following the procedure} described in {Ref.} \cite{Ginibre}.  {First, w}e construct {a} $d\times r$ {complex Ginibre} matrix $G$, 
\begin{align}
G=
    \begin{pmatrix}
     g_{11}&\cdots &g_{1r}\\
     &\vdots&\\
     g_{d1}&\cdots&{g}_{dr}
    \end{pmatrix},
\end{align}
{whose} elements  $g_{ij}$ are independently drawn from a standard complex normal distribution,
\begin{align}
   {g_{ij}=\mathcal{N}(0,1)+i\,\mathcal{N}(0,1),}
\end{align}
{where $\mathcal{N}(0,1)$ represents a real-valued Gaussian distribution} with zero mean and unit variance. {Using $G$, we construct the density matrix $\rho$ of a target random state as}
\begin{align}
    {\rho=\frac{GG^\dag}{\mathrm{Tr}[GG^\dag]}.}
\end{align}
This {procedure} ensures a {randomly generated} valid quantum state, with unit trace and positive semi-definiteness{, where the matrix rank is 
$r=1$ for a pure state or $r\geq 2$ for a mixed state.}

\subsection{{Simulation setup and procedure}}
\label{methods}
{The setup of the numerical simulation to test the accuracy of CS-QST is as follows.} First, the target density matrix {$\rho$} is generated using the Ginibre ensemble, as introduced in Sec.~\ref{Ginibre ensemble}. {For simplicity and to focus solely on the impact of the choice of measurement basis, we consider only} pure states {with rank $r=1$. To mimic state preparation errors, we apply} depolarizing noise at a level of 5\%, as described by
 \begin{align}
     \tilde{\rho}=(1-0.05)\rho+\frac{0.05}{d}I,
 \end{align}
{where $I$ is the $d$-dimensional identity matrix.}

{We simulate} expectation value measurements using $m$ randomly selected {measurement operators, sampled from $\{ B_a\}_{a=0}^{d^2-1}$. To model} measurement noise, {we add a noise term $\sigma$, sampled from a normal distribution $\mathcal{N}(0,0.1/d)$. The noisy expectation values are} given by
\begin{align}
{\langle B_a \rangle_{\mathrm{n}} = \mathrm{Tr}[\tilde\rho B_a]+\sigma,}
\end{align}
{where the subscript ``n'' distinguishes from the exact expectation value. In principle, using all the expectation values $\langle {B_a}\rangle_{\mathrm{n}}$ and the corresponding measurement operators $B_a$, one could find the best approximation to the reconstructed matrix $\bm{M}$ as}
\begin{align} 
\bm{M} = \sum_{a=0}^{d^2-1}  {\langle {B}_a \rangle_{\mathrm{n}} } B_a, 
\end{align}
{which coincides with the target density matrix $\rho$ except for the effects of preparation and measurement noise. In CS-QST, to reduce the measurement cost, we instead construct a sample matrix }
\begin{align} 
{\mathcal{P}_\Omega(\bm{M}) = \sum_{a\in \Omega} { \langle {B}_a \rangle_{\mathrm{n}} } B_a, }
\label{projomega}
\end{align}
{where $\Omega\subset \{0,1,\dots, d^2-1\}$ is a randomly selected subset of size $m$, and $\mathcal{P}_\Omega$ denotes the orthogonal projection onto the {linear} subspace spanned by { $\{B_a\}_{a\in \Omega}$}.} 

Starting from the sample matrix $\mathcal{P}_\Omega(\bm{M})$, we apply the SVT algorithm~\cite{SVT} to reconstruct the density matrix. A brief review of the SVT algorithm, along with the specific parameters used in our simulations, is provided in Appendix~\ref{SVT}. The {output} matrix is denoted as $\bm{X}$; {to obtain a \textit{bona fide} density matrix, we additionally operate a trace normalization and arrive at the final reconstructed density matrix $\tilde{\bm{X}}$.}  Its accuracy in reproducing the original density matrix is evaluated using two standard metrics, fidelity and trace distance, defined as follows:
  \begin{align}
      &F(\rho,{\tilde{\bm{X}}})=\frac{{\mathrm{Tr}}(\sqrt{\sqrt{\rho}{\tilde{\bm{X}}}\sqrt{\rho}})^2}{{\mathrm{Tr}(\rho)\mathrm{Tr}({\tilde{\bm{X}}})}}{,}\\
      &D(\rho,{\tilde{\bm{X}}})=\frac{1}{2}\mathrm{Tr}|\rho-{\tilde{\bm{X}}}|{.}
  \end{align}
{These quantities are used to quantitatively assess the performance of the CS-QST reconstruction.

We repeat the above procedure 50 times and perform statistical {analysis on the results}. First, we compute the trace of each reconstructed matrix {$\tilde{\bm{X}}$}{,} and {only} consider {those} satisfying $0\leq \mathrm{Tr}[\bm{X}]<2$ as {valid} CS-QST {outcomes} {(in practice, in our calculation, invalid outcomes occur very rarely and for very small $m/d^2$ \footnote{Typically they are less than $4\%$ for the lowest one or two values of $m/d^2$ in the simulations presented in Sec.~\ref{results}, except for one instance in which they are $8\%$, namely GGM basis, $k=9$ two-qudit states, $m/d^2=0.02$. })}. The mean value{s are} then computed using these selected samples.  Error bars are defined as the range of $\pm 1$ standard deviation{, calculated} as follows:
\begin{align}
\begin{split}
     &S_\mathrm{F}=\sqrt{\frac{1}{{N_{\rm reps}}-1}\sum_{{\mu}=1}^{N_{\rm reps}}({F}(\rho,{\tilde{\bm{X}}})_{\mu}-{\bar{F}(\rho,{\tilde{\bm{X}}})})}{,} \\\\
    &S_\mathrm{D}=\sqrt{\frac{1}{{N_{\rm reps}}-1}\sum_{{\mu}=1}^{N_{\rm reps}}({D}(\rho,{\tilde{\bm{X}}})_{\mu}-{\bar{D}(\rho,{\tilde{\bm{X}}})})}{,} 
\end{split}
\end{align}
where {$N_{\rm reps}$} is the number of {valid} repetitions, and {$\bar{F}, \bar{D}$ denote} the {sample means of the} fidelity ${F(\rho,{\tilde{\bm{X}}})}$ {and the} trace distance ${D(\rho,{\tilde{\bm{X}}})}$, {respectively}. {The subscript $\mu$ indicates the value obtained in the $\mu$-th repetition. {Note that $N_{\rm reps}$ remains nearly 50,
as most samples are valid. }}

\subsection{{Main} results}
\label{results}
\begin{figure*}[htbp]
    \includegraphics[keepaspectratio, scale=0.9]{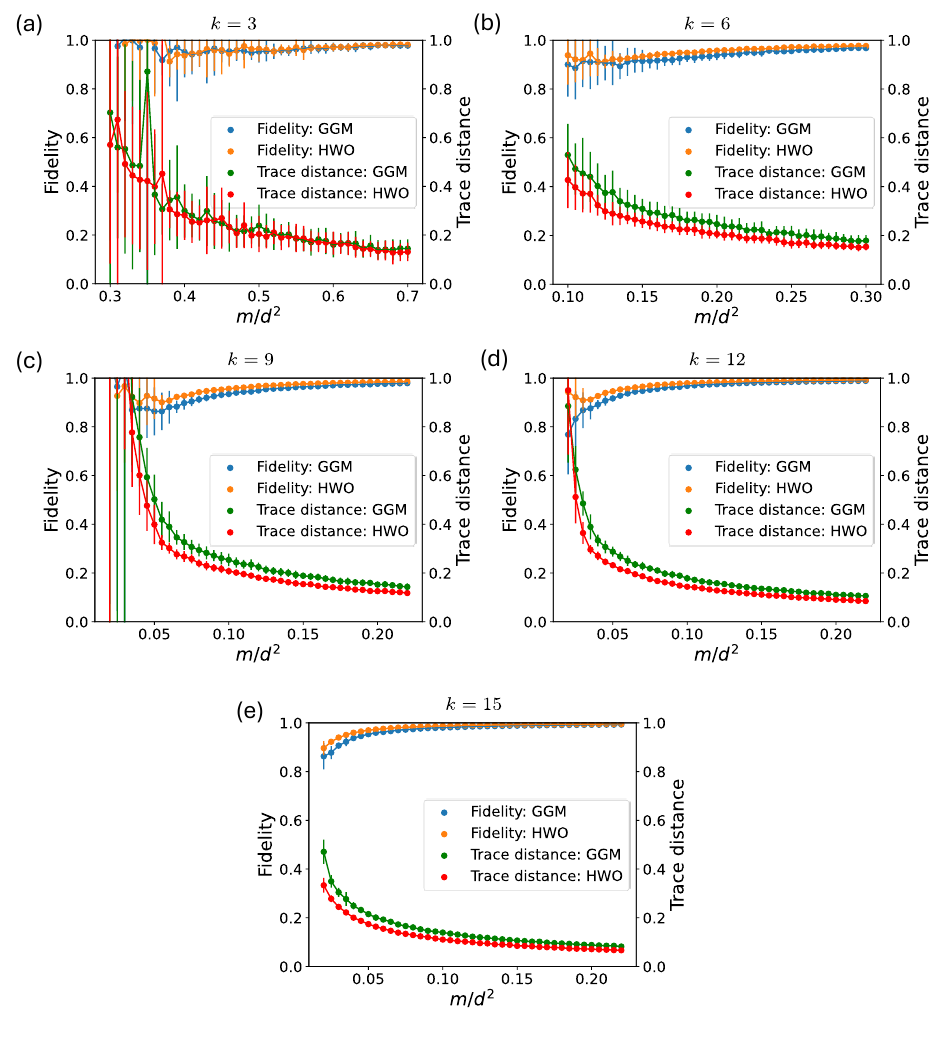}
     \caption{{Fidelity $F(\rho,{\tilde{\bm{X}}})$ and trace distance $D(\rho,{\tilde{\bm{X}}})$ between the target two-qudit state $\rho$ and the matrix ${\tilde{\bm{X}}}$ obtained via CS-QST, for qudit dimensions (a) $k=3$, (b) 6, (c) 9, (d) 12, and (e) 15. The horizontal axis represents the number of randomly selected measurement operators $m$, scaled by the total number of basis elements $d^2=k^4$. }}
     \label{ratios}
  \end{figure*}
  
\subsubsection{{Two-qudit states}}
\label{Impact}
{Figures}~\ref{ratios}(a-e) show the efficiency of {CS-QST for {an ensemble of random pure states} of two qudits ($N=2$) using} two different bases{,} as {a function of} the number $m$ of randomly selected measurement operators, {for qudit dimensions $k=3,6,9,12$, and $15$, respectively}. The results provide insights into how the choice of basis affects reconstruction accuracy under varying conditions, particularly {with respect} to the number of measurements. 

For $k=3$, we observe no significant difference in reconstruction performance between the HWO and GGM bases. Both bases {yield} large { fluctuations in fidelity and trace distance (especially for $m/d^2\lesssim 0.5$)}, and {neither shows a} clear advantage. {These findings suggest} that, in lower-dimensional systems, the choice of basis may not be a {critical} factor in the efficiency of CS-QST. 
In contrast, for larger dimensions ($k=6, 9, 12, 15$), a clear trend emerges in favor of the HWO basis. In these cases, the HWO basis tends to reconstruct the density matrix more efficiently than the GGM basis{,} particularly when the number of measurements $m$ is small. {Additionally,} under such conditions, the GGM basis exhibits {somewhat} larger error bars, indicating {greater} variability in reconstruction accuracy. {This} behavior can be attributed to the nature of the randomly selected measurement operators. When $m$ is small, the likelihood of selecting a suboptimal measurement set---one that is less compatible with the target state---increases, leading to larger {reconstruction} errors. {This effect is more pronounced for the GGM basis, as its coherence varies across operators, as discussed in Sec.~\ref{sec_coherence}. } These results suggest that the HWO basis may offer a more stable reconstruction framework under {conditions of limited measurement numbers.}

As $m$ increases, the error bars for both bases {become negligible, and the difference in the reconstruction accuracy between them simultaneously diminishes. Although the HWO basis continues to provide} a more efficient reconstruction of the density matrix compared to the GGM basis, the overall difference between the two  becomes smaller. {This convergence in reconstruction accuracy can be attributed to the fact that, as the number of measurement operators increases, the statistical effects of suboptimal operator selection are mitigated.}

\begin{figure*}[tb]
    \centering
    \includegraphics[width=0.9\linewidth]{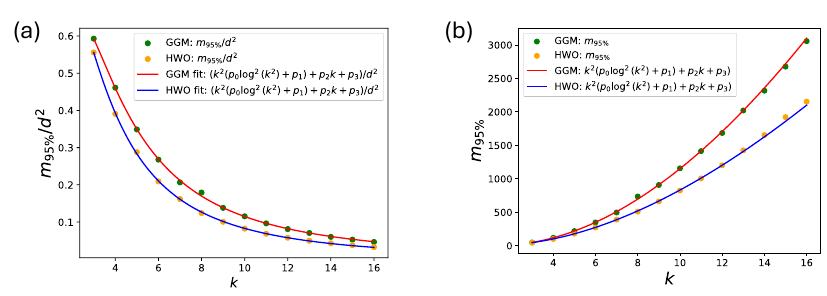}
    \caption{{(a) The relative measurement fraction $m_{95\%}/d^2$, obtained by normalizing the number of measurements $m_{95\%}$ required to achieve a fidelity of at least 95\% in CS-QST of two-qudit states by the total number of measurement operators $d^2=k^4$, is plotted as a function of the qudit dimension $k$, along with the corresponding fitted curves. The fitted parameters for the GGM basis are $p_0=-0.117,~p_1=18.865,~p_2=-53.044$, and $p_3=42.449$; for the HWO basis, $p_0=-0.123,~ p_1=13.897,~ p_2=-31.882$, and $p_3=20.894$. (b) The same  data as in (a), but {with} the vertical axis rescaled by  $d^2$, showing the number of measurements $m_{95\%}$ }}
    \label{ratio_and_num}
\end{figure*}

Next, {to compare the efficiency  in more detail}, we investigate the number of measurements $m$ required to achieve a fidelity of at least 95\%. {W}e impose the following conditions: {(i) the mean fidelity minus one standard deviation remains above 95\%,}
and (ii) further increasing $m$ does not {cause the fidelity to fall} below 95\%. The smallest {value of} $m$ that satisfies {both conditions, referred to as $m_{95\%}$}, is {identified} as the required number of measurements. The results are presented in Figs.~\ref{ratio_and_num}(a) and~\ref{ratio_and_num}(b).

In Fig.~\ref{ratio_and_num}(a), {we first analyze the relative measurement cost $m_{95\%}/d^2$, which provides an important perspective by quantifying the number of measurements required relative to the total operator space dimension.}
The functional form of the required  {relative measurement cost $m_{95\%}/d^2$} for the GGM and HWO bases, obtained by least-squares fitting,} can be expressed as follows:
{
\begin{align}
{m_{95\%}^{\mathrm{GGM}}/d^2= {[} k^2(p_0\log^2{(k^2)}+p_1)+p_2k+p_3} {]} /d^2
\label{gelfit}
\end{align}
{with $p_0=-0.117,~p_1=18.865$, $p_2=-53.044$, and $p_3=42.449$;}
\begin{align}    
{m_{95\%}^{\mathrm{HW}}/d^2={[} k^2(p_0\log^2{(k^2)}+p_1)+p_2k+p_3} {]} /d^2
    \label{HWfit}
\end{align}
{with $p_0=-0.123,~p_1=13.897$, $p_2=-31.882$, and $p_3=20.894$}}.  {The above forms were selected among several candidate fitting models based} on the Akaike Information Criterion (AIC)~\cite{AIC}, which quantitatively evaluates the trade-off between goodness of fit and model complexity (number of fitting parameters). Other candidate {functions} and their corresponding AIC values are {reported} in {A}ppendix~\ref{apendAIC}.  {These results are broadly consistent with the expected scaling law $m\sim O(rd\log^2d)$ in compressed sensing~\cite{CS}, as both fitted forms take the structure ${ {[} k^2\left(p_0\log^2(k^2) + p_1\right)+p_2k+p_3 {]} /d^2}$. However, the fitted values of {$p_1$} are significantly larger than those of {$p_0$}, indicating that the scaling is closer to  {$m_{95\%}/d^2\sim k^2/d^2=1/k^2$} in practice, at least in the range of dimensions at hand. Moreover, {the coefficients in the GGM fitting function, particularly $p_1,~p_2$, and $p_3$ are significantly larger than those in the HWO case. This} may {be a consequence of} the coherence difference between diagonal and off-diagonal elements in the GGM basis, as discussed in Sec.\ref{sec_coherence}.}

 We next report the absolute number of measurements $m_{95\%}$ required to reach a fidelity of at least 95\% in Figure~\ref{ratio_and_num}(b). When the qudit dimension $k$ is small ($k\lesssim 5$), the required number of measurements {appears} comparable between the two bases. However, as $k$ increases, a clear discrepancy begins to emerge: the HWO basis consistently requires fewer measurements than the GGM basis, with the gap widening for larger $k$ {In fact, the growth rate of the gap is approximately $k^2$, as inferred from the fitted curve.} These results indicate that the HWO basis {offers} a more efficient approach {to} quantum state reconstruction  in higher-dimensional qudit systems.
This is consistent with expectations based on the coherence structure of the measurement bases, described in Sec.~\ref{sec_coherence}. The HWO basis exhibits nearly constant minimum coherence $\nu_{\rm min}$ regardless of the qudit dimension $k$, whereas the coherence for the GGM basis increases as $(k-1)^N$. This increasing coherence makes the GGM basis less favorable in high-dimensional systems, where the HWO basis provides a more stable and efficient reconstruction framework.

\subsubsection{{Fixed Hilbert space dimension}}
\begin{figure}
    \centering
    \includegraphics[width=1.1\linewidth]{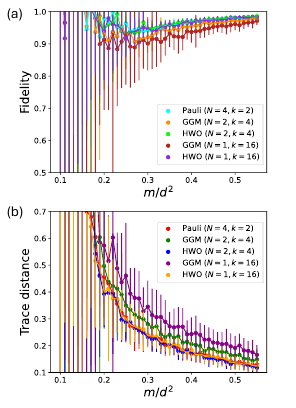}
    \caption{{Comparison of CS-QST performance for quantum systems with the same Hilbert space dimension $d=16$, represented using different combinations of qubit and qudit systems. Panel (a) shows the fidelity, and panel (b) the trace distance, as functions of the number of randomly selected measurement operators $m$, normalized by the total number of operators $d^2$.}}
    \label{compare}
\end{figure}


Finally, we examine how the choice of {quantum information unit}---{qubit} or qudit---affects the efficiency of CS-QST when the Hilbert space dimension $d$ is fixed. By varying 
$N$ and $k$ while keeping $d=k^N$ constant, we compare the relative advantages of different representations. 
Specifically, we {focus on the case of } a density matrix with $d=16${,} and analyze the following {three configurations:} (i) a four-qubit system using the Pauli basis{;} (ii) a two-qudit system {with} $k=4$, using {either the GGM or} HWO basis; and (iii) a single-qudit system {with} $k=16$, using {either the GGM or} HWO basis. 

{In Figs.~\ref{compare}(a) and~\ref{compare}(b), we show the fidelity and trace distance, respectively, as functions of the normalized number of measurements $m/d^2$ for the three configurations introduced above. For the two- and single-qudit cases, both the GGM and HWO bases are considered, resulting in five total curves. }
From these results, we observe that {the efficiency of density matrix reconstruction is comparable} when using qubits or the HWO basis for qudits. {In contrast, when the GGM basis is employed,} the reconstruction efficiency deteriorates as the qudit {dimension $k$} increases.

This difference in reconstruction efficiency can be understood again in terms of the coherence structure of the measurement bases. In the case of the GGM basis, the minimum coherence $\nu_{\rm min}$ grows as $(k-1)^N$, as shown in Sec.~\ref{sec_coherence}, indicating that the measurement operators become less favorable for compressed sensing when increasing $k$ at fixed $d$. In contrast, the HWO basis maintains a nearly constant $\nu_{\rm min}$ {versus} $k$, contributing to its robustness even in high-dimensional systems. Notably, for $k=4$, the HWO basis yields $\nu_{\rm min}=1$, identical to the Pauli basis for qubits. Even for $k=16$, the value remains low as long as the number of qudits $N$ is small, with $\nu_{\rm min}=2^N$ (see Table \ref{tableHW}). This favorable coherence behavior helps to explain the consistently higher efficiency observed with the HWO basis in our numerical simulations.
\color{black}

\section{Summary}
\label{summary}

We have numerically studied compressed sensing quantum state tomography (CS-QST) for qudit systems, focusing on how the choice of measurement basis affects reconstruction efficiency. In particular, we compared {the performance of} the generalized Gell-Mann (GGM) and Heisenberg-Weyl {observable} bases, {which differ qualitatively in terms of their coherence properties and how they relate to the CS-QCS theorem of Ref.~\cite{cohe}},  in quantum state reconstruction of two-qudit ($N = 2$) systems. Our simulations showed that while both bases can achieve high-fidelity reconstruction, the HWO basis consistently outperforms the GGM basis, especially as the qudit dimension $k$ increases. We also analyzed how the number of measurements required for accurate reconstruction scales with $k$.

Furthermore, we compared the performance of CS-QST across different configurations that realize the same Hilbert space dimension $d = k^N$ using either qubits or qudits. The results show that, when using the HWO basis, qudit-based tomography achieves reconstruction performance comparable to qubit-based tomography, even at large $k$. In contrast, reconstructions using the GGM basis exhibit a gradual decline in performance as $k$ increases.

These differences can be consistently understood in terms of the coherence structure of the measurement bases. For the GGM basis, the minimum coherence $\nu_{\rm min}$, defined in Eq.~\ref{mincoherence}, scales as $(k - 1)^N$, making it increasingly incompatible with compressed sensing as the qudit level $k$ increases. In contrast, the HWO basis yields $\nu_{\rm min} \approx 2^N$, which remains {essentially} constant when $k$ is varied at fixed Hilbert space dimension $d = k^N$. Since increasing $k$ reduces $N$, the coherence remains within a manageable range even for large $k$, further supporting the scalability of the HWO basis in high-dimensional CS-QST.

Our findings provide quantitative insights into the role of basis choice in CS-QST and highlight the HWO basis as a promising approach for scalable and efficient quantum state reconstruction in high-dimensional qudit systems. Nevertheless, we emphasize that the GGM basis {can still provide a} sufficiently accurate reconstruction in practice. Therefore, the choice between GGM and HWO bases {(and, in principle, other bases with various coherence properties)} should be guided by experimental constraints {and goals}, such as the ease of implementing measurements {of certain observables} and the fidelity requirements of the specific application.

\begin{acknowledgments}
{We would like to thank Y.~Miyazaki for useful discussions. The work of was supported by JSPS KAKENHI Grant Nos.~21H05185 (D.Y., G.M.), 23K25830 (D.Y.), 24K06890 (D.Y.), and JST PRESTO Grant Nos.~JPMJPR2118 (D.Y.) and JPMJPR245D (D.Y.).} 
\end{acknowledgments}

\section{appendix}
\subsection{{GGM and HW-observable operators for $k=4$}}
\label{aappendixGGM}
{For convenience, we present here the matrix representations of the GGM and HW-observable operators for $k=4$. The GGM matrices for $k = 4$ are categorized into three types:} 
\begin{description}
    \item[i)] Six symmetric GGM matrices 
    \begin{align*}
    \begin{split}
        &\Lambda^{12}_s=
        \begin{pmatrix}
            0&1&0&0\\1&0&0&0\\0&0&0&0\\0&0&0&0
        \end{pmatrix}
        ,\hspace{5pt}
        \Lambda^{13}_s=
        \begin{pmatrix}
            0&0&1&0\\0&0&0&0\\1&0&0&0\\0&0&0&0
        \end{pmatrix}
        \\
        &\Lambda^{14}_s=
        \begin{pmatrix}
            0&0&0&1\\0&0&0&0\\0&0&0&0\\1&0&0&0
        \end{pmatrix},\hspace{5pt}
        \Lambda^{23}_s=
        \begin{pmatrix}
            0&0&0&0\\0&0&1&0\\0&1&0&0\\0&0&0&0
        \end{pmatrix}
        \\
        &\Lambda^{24}_s=
        \begin{pmatrix}
            0&0&0&0\\0&0&0&1\\0&0&0&0\\0&1&0&0
        \end{pmatrix}
        ,\hspace{5pt}
        \Lambda^{34}_s=
        \begin{pmatrix}
            0&0&0&0\\0&0&0&0\\0&0&0&1\\0&0&1&0
        \end{pmatrix}\\
        \end{split}
    \end{align*}
    
    \item[ii)] Six anti-symmetric GGM matrices 
    \begin{align*}
    \begin{split}
        &\Lambda^{12}_a=
        \begin{pmatrix}
            0&-i&0&0\\i&0&0&0\\0&0&0&0\\0&0&0&0
        \end{pmatrix}
        ,\hspace{5pt}
        \Lambda^{13}_a=
        \begin{pmatrix}
            0&0&-i&0\\0&0&0&0\\i&0&0&0\\0&0&0&0
        \end{pmatrix}\\
        &\Lambda^{14}_a=
        \begin{pmatrix}
            0&0&0&-i\\0&0&0&0\\0&0&0&0\\i&0&0&0
        \end{pmatrix}
        ,\hspace{5pt}
        \Lambda^{23}_a=
        \begin{pmatrix}
            0&0&0&0\\0&0&-i&0\\0&i&0&0\\0&0&0&0
        \end{pmatrix}
        \\
        &\Lambda^{24}_a=
        \begin{pmatrix}
            0&0&0&0\\0&0&0&-i\\0&0&0&0\\0&i&0&0
        \end{pmatrix}
        ,\hspace{5pt}
        \Lambda^{34}_a=
        \begin{pmatrix}
            0&0&0&0\\0&0&0&0\\0&0&0&-i\\0&0&i&0
        \end{pmatrix}\\
        \end{split}
    \end{align*}
    
    \item[iii)] Three diagonal GGM matrices 
\begin{align*}
    \begin{split}
        &\Lambda_d^1=
        \begin{pmatrix}
            1&0&0&0\\0&-1&0&0\\0&0&0&0\\0&0&0&0
        \end{pmatrix},\quad
          \Lambda_d^2=\frac{1}{\sqrt{3}}
        \begin{pmatrix}
            1&0&0&0\\0&1&0&0\\0&0&-2&0\\0&0&0&0
        \end{pmatrix}\\
        &\Lambda_d^3=\frac{1}{\sqrt{6}}
        \begin{pmatrix}
            1&0&0&0\\0&1&0&0\\0&0&1&0\\0&0&0&-3
        \end{pmatrix}
    \end{split}
\end{align*}
\end{description}

The HW-observable matrices for $k=4$, including the identity matrix $W(0,0)$ are listed below. Here, we use $\chi = (1 + i)/2$ as defined in the main text.
\begin{widetext}
\begin{align*}
\begin{split}
    & W(0,0)=
    \begin{pmatrix}
        1&0&0&0\\0&1&0&0\\0&0&1&0\\0&0&0&1
    \end{pmatrix},\quad
    W(0,1)=
    \begin{pmatrix}
        0&\chi^*&0&\chi\\\chi&0&\chi^*&0\\0&\chi&0&\chi^*\\\chi^*&0&\chi&0
    \end{pmatrix} , \quad
    W(0,2)=
    \begin{pmatrix}
        0&0&1&0\\0&0&0&1\\1&0&0&0\\0&1&0&0
    \end{pmatrix} \\
   & W(0,3)=
    \begin{pmatrix}
        0&\chi&0&\chi^*\\\chi^*&0&\chi&0\\0&\chi^*&0&\chi\\\chi&0&\chi^*&0
    \end{pmatrix}, \quad
    W(1,0)=
    \begin{pmatrix}
        1&0&0&0\\0&-1&0&0\\0&0&-1&0\\0&0&0&1
    \end{pmatrix}, \quad
    W(1,1)=\frac{1}{\sqrt{2}}
    \begin{pmatrix}
        0&-i&0&1\\i&0&-1&0\\0&-1&0&i\\1&0&-i&0
    \end{pmatrix}, \\
    & W(1,2)=
    \begin{pmatrix}
        0&0&-i&0\\0&0&0&i\\i&0&0&0\\0&-i&0&0
    \end{pmatrix}, \quad
    W(1,3)=\frac{1}{\sqrt{2}}
    \begin{pmatrix}
        0&-i&0&-1\\i&0&1&0\\0&1&0&i\\-1&0&-i&0
    \end{pmatrix},\quad
    W(2,0)=
    \begin{pmatrix}
        1&0&0&0\\0&-1&0&0\\0&0&1&0\\0&0&0&-1
    \end{pmatrix}, \\
    & W(2,1)=
    \begin{pmatrix}
        0&-\chi&0&\chi^*\\-\chi^*&0&\chi&0\\0&\chi^*&0&-\chi\\\chi&0&-\chi^*&0
    \end{pmatrix},\quad 
    W(2,2)=
    \begin{pmatrix}
        0&0&-1&0\\0&0&0&1\\-1&0&0&0\\0&1&0&0
    \end{pmatrix}, \quad
    W(2,3)=
    \begin{pmatrix}
        0&-\chi^*&0&\chi\\-\chi&0&\chi^*&0\\0&\chi&0&-\chi^*\\\chi^*&0&-\chi&0
    \end{pmatrix}, \\
    & W(3,0)=
    \begin{pmatrix}
        1&0&0&0\\0&1&0&0\\0&0&-1&0\\0&0&0&-1
    \end{pmatrix},\quad 
    W(3,1)=\frac{1}{\sqrt{2}}
    \begin{pmatrix}
        0&-1&0&-i\\-1&0&-i&0\\0&i&0&1\\i&0&1&0
    \end{pmatrix},\quad 
    W(3,2)=
    \begin{pmatrix}
        0&0&i&0\\0&0&0&i\\-i&0&0&0\\0&-i&0&0
    \end{pmatrix}, \\
    & W(3,3)=\frac{1}{\sqrt{2}}
    \begin{pmatrix}
        0&1&0&-i\\1&0&-i&0\\0&i&0&-1\\i&0&-1&0
    \end{pmatrix}.
\end{split}
\end{align*}
\end{widetext}

\subsection{Minimum coherence of the HWO basis when $k$ is a multiple of 8}\label{appendix2}

Here, we provide an explanation for why the minimum coherence $\nu_{\rm min}$ of the HW-observable basis becomes exactly $2^N$ when the qudit dimension $k$ is a multiple of 8. 

The HW-observable matrices defined in Eqs.~\eqref{HWop1} and~\eqref{HWop2} share the same set of eigenvalues (except for $W(0,0)$). As can be easily seen, for example in $W(1,0)$, the eigenvalues are given by
\begin{eqnarray}
    \epsilon_n&=&\chi e^{i2\pi (n-1)/k}+\chi^\ast e^{-i2\pi (n-1)/k},
\end{eqnarray}
where $n = 1, 2, \dots, k$ and $\chi = (1+i)/2$ as used in this study. The minimum coherence defined in Eq.~\eqref{mincoherence} becomes 
\begin{eqnarray}
    \nu_{\rm min}&=&\left(\max_n \epsilon_n^2\right)^N\nonumber
\\
    &=&\left(\max_n \left(1-\sin \frac{4\pi(n-1)}{k}\right)\right)^N.\label{mincoherenceHW}
\end{eqnarray}
This expression reaches its maximum value of $2^N$ if and only if $k$ is a multiple of 8, because the sine term $\sin(4\pi(n-1)/k)$ can attain $-1$ when $4\pi(n-1)/k \equiv 3\pi/2 \ (\mathrm{mod}\ 2\pi)$. In such cases, there are two values of $n$ for which $\epsilon_n^2 = 2$, and thus $\nu_{\rm min} = 2^N$. It is also worth noting that Eq.~\eqref{mincoherenceHW} explains why $\nu_{\rm min}$ takes the exact value of 1 for $k = 4$.
\color{black}

\subsection{SVT {algorithm}}\label{SVT}
Here, we {briefly review} the SVT algorithm \cite{SVT} {and summarize} the parameter {settings} used in this study. 

{Starting from} the sample matrix {$\mathcal{P}_{\Omega}(\bm{M})$}, obtained from expectation value measurements {over a} randomly {chosen subset of} operators (see Eq.~\ref{projomega} and the explanation thereafter), we {define the initial matrix $\bm{Y}^0$ as}
\begin{align}
    \bm{Y}^0={t_{0}}\delta \mathcal{P}_{\Omega}(\bm{M}).
\end{align}
{Here,}  ${t_{0}}$ is an {initial} iteration {counter} and $\delta$ is {the step size} given by
 \begin{align}
      \delta=0.1\times k^4/m \hspace{1mm}(=0.1\times d^2/m),
 \end{align}
{where} $k$ is the qudit dimension {and $m$ is the number of selected measurement operators}.
 
 The integer {$t_{0}$ is} chosen {such that}:
\begin{align}
\label{t_0}
    \frac{\tau}{\delta\|\mathcal{P}_\Omega (\bm{M})\|_2}\in(t_{0}-1,t_0\rbrack.
\end{align}

{Although the SVT} algorithm {is typically} iterated from {$t= 1$ to $t_{\rm max}$}, it is known that, {by selecting $t_{0}$ as above, early iterations can be effectively skipped~\cite{SVT}. In this study, we set $t_{\rm max}=2\times t_{0}$ to maintain consistency in the number of SVT steps across different samples and ensure uniform numerical conditions.}

The SVT algorithm then proceeds iteratively as follows. At each iteration $t$, the matrix $\mathbf{X}^{(t)}$ is updated by applying a singular value thresholding operator to the $\mathbf{Y}^{(t-1)}$ {of the previous step}:
\begin{align}
\mathbf{X}^{(t)} = \mathcal{D}_\tau(\mathbf{Y}^{(t-1)}),
\end{align}
where $\mathcal{D}_\tau(\cdot)$ denotes the soft-thresholding operation~\cite{SVT} on the singular values with threshold $\tau$.

The subsequent $\mathbf{Y}^{(t)}$ is then updated based on the discrepancy between the observed data and the projection of the current estimate:
\begin{align}
\mathbf{Y}^{(t)} = \mathbf{Y}^{(t-1)} + \delta \, \mathcal{P}_\Omega(\mathbf{M} - \mathbf{X}^{(t)}).
\end{align}

These steps are repeated until the iteration count reaches $t_{\mathrm{max}}$ or the change between successive iterations falls below the convergence tolerance $\epsilon$. In this study, we {fix} $\tau = 5$ for the threshold, $l = 2$ as the increment parameter in the original SVT method, and the convergence tolerance is set to $\epsilon = 10^{-7}$.
\color{black}

\subsection{{Model selection using AIC}}
\label{apendAIC}

{In order to determine an appropriate model for fitting the numerical data {in Fig.~\ref{ratio_and_num}}, we use} the Akaike Information Criterion (AIC)~\cite{AIC}, {a statistical measure that evaluates the relative quality of models by balancing} the goodness of fit {and} model complexity. {The AIC} is defined as
\begin{align}
\label{AICfo}
    {\rm AIC}=-2\log {\hat{\mathcal{L}}} + 2{n_{\rm p}}
\end{align}
where ${\hat{\mathcal{L}}}$ denotes the maximum value of the likelihood function, and ${n_{\rm p}}$ is the number of {fitting} parameters. 
{A lower AIC value indicates a better model. 

We fit the data points of $m_{95\%}(k)$ in Fig.~\ref{ratio_and_num}(a) using several candidate models, including polynomials and logarithmic-polynomial hybrids. The {value of the fitting function} at each point is denoted by $\hat{m}_{95\%}(k)$.}

The residual sum of squares (RSS) is computed as
\begin{align}
   \mathrm{RSS}=\sum_{k}(m_{95\%}(k)-\hat{m}_{95\%}(k))^2.
\end{align}
The mean squared error (MSE), used to estimate the variance of normally distributed residuals, is
\begin{align}
    \sigma^2=\frac{1}{N_{\rm data}}\mathrm{RSS},
\end{align}
where $N_{\rm data}= 14$ is the number of data points.
Assuming normally distributed residuals, the log-likelihood becomes
\begin{align*}
    \log \hat{\mathcal{L}} = \sum_{k} \left[ -\frac{1}{2} \log(2\pi\sigma^2) - \frac{(m_{95\%}(k)-\hat{m}_{95\%}(k))^2}{2\sigma^2} \right].
\end{align*}
This expression is used to calculate the AIC for each model. 

The results of applying the AIC to various {fitting} curves using the methodology described above are presented in Table~\ref{AICtable}. Among the candidate {models}, the one with the lowest AIC value was selected as the optimal fitting model. The corresponding {fitting} curves are shown in Fig.~\ref{ratio_and_num}(a), while the same curves multiplied by $d^2$ are presented in Fig.~\ref{ratio_and_num}(b).

\color{black}

\begin{table}[htb]
 \caption{{AIC values for each candidate fitting {model}. The {model} with the lowest AIC in each column (GGM or HWO) is considered the best fit.}}
 \begin{center}
 \begin{tabular}{c|cc}
 \hline\hline
     Fitting formula&GGM&HW\\
     \hline
     {$(p_0k^2+p_1k+p_2)/d^2$}&~{-113.6}~&~{-124.0~}\\
     \hline
     {${[}k^2(p_0\log^2k^2+p_1){]}/d^2$}&{~-66.7}&~{-75.7~}\\
     \hline
     {${[} k^2(p_0\log^2k^2+p_1)+p_2k{]} /d^2$}&~{-112.0}~&~{-127.9}~\\
     \hline
     {~${[} k^2(p_0\log^2k^2+p_1)+p_2k+p_3{]} /d^2$}~~&{~-114.7}~&~{-130.6}~\\
     \hline\hline
\end{tabular}
\label{AICtable}
\end{center}
\end{table}

\bibliography{apssamp}

\end{document}